\documentclass[aps,prl,twocolumn,superscriptaddress]{revtex4}

\usepackage{graphicx}

\begin{document}

\preprint{}

\title{Uniform Mixing of Antiferromagnetism and High-$T_{\rm c}$ Superconductivity in Electron-doped Layers in Four-layered Ba$_2$Ca$_3$Cu$_4$O$_8$F$_2$ : A New Phenomenon in an Electron Underdoped Regime}

\author{S. Shimizu}
\email[]{e-mail: shimizu@nmr.mp.es.osaka-u.ac.jp}
\author{H. Mukuda}
\author{Y. Kitaoka}
\affiliation{Department of Materials Engineering Science, Osaka University, Osaka 560-8531, Japan }

\author{A. Iyo}
\author{Y. Tanaka}
\author{Y. Kodama}
\affiliation{National Institute of Advanced Industrial Science and Technology (AIST), Umezono, Tsukuba 305-8568, Japan}

\author{K. Tokiwa}
\author{T. Watanabe}
\affiliation{Department of Applied Electronics, Tokyo University of Science, Noda, Chiba 278-8510, Japan}

\date{\today}

\begin{abstract}
We report $^{63,65}$Cu- and $^{19}$F-NMR studies on a four-layered high-temperature superconductor Ba$_2$Ca$_3$Cu$_4$O$_8$F$_2$(0234F(2.0)) with apical fluorine (F$^{-1}$), an {\it undoped} 55 K-superconductor with a {\it nominal Cu$^{2+}$ valence} on average.  We reveal that this compound exhibits the antiferromagnetism (AFM) with a Neel temperature $T_{\rm N}$=100 K despite being a $T_{\rm c}=$ 55 K-superconductor. Through a comparison with a related tri-layered cuprate Ba$_2$Ca$_2$Cu$_3$O$_6$F$_2$ (0223F(2.0)), it is demonstrated that electrons are transferred from the inner plane (IP) to the outer plane (OP) in 0234F(2.0) and 0223F(2.0), confirming the {\it self-doped} high-temperature superconductivity (HTSC) having electron and hole doping in a single compound. Remarlably, uniform mixing of AFM and HTSC takes place in both the {\it electron-doped} OPs and the {\it hole-doped} IPs in 0234F(2.0). 
\end{abstract}


\maketitle

The origin of high-temperature superconductivity (HTSC) is still not well understood despite 20 year's intensive research. In particular, a possible interplay between antiferromagnetism (AFM) and HTSC remains one of the most interesting problems \cite{Anderson,Inaba,Zhang,Himeda,Sidis,Lake,TKLee,Demler}. All HTSC compounds share a layered structure made up of one or more copper-oxygen (CuO$_2$) planes. 
Three or more layered cuprates consist of inequivalent types of CuO$_2$ layers, an outer CuO$_2$ plane (OP) in a five-fold pyramidal coordination and an inner plane (IP) in a four-fold square one. 
Previous $^{63}$Cu-NMR measurements \cite{Tokunaga,Kotegawa2001,Kotegawa2004,Mukuda} revealed that the IP has less holes than the OP and each flat CuO$_2$ plane is homogeneously doped, which brings about several unique magnetic and superconducting (SC) properties inherent to multi-layered systems.  An extreme example is five-layered HgBa$_2$Ca$_4$Cu$_5$O$_{y}$(Hg-1245(optimally)) \cite{Kotegawa2004} in which the optimally-doped OPs are responsible for the HTSC with $T_{\rm c}=108$ K and the three underdoped IPs are AFM metals with moments of 0.3$\sim$0.37 $\mu_{\rm B}$ below $T_{\rm N}=60$ K. By reducing the hole density, $T_{\rm c}$ goes down to 72 K and $T_{\rm N}$ up to 290 K for the underdoped Hg-1245(UD) \cite{Mukuda}. Eventually, the hole-doped OPs in Hg-1245(UD) exhibit uniform mixing of AFM and HTSC with an AFM moment of $M({\rm OP})=0.1$ $\mu_{\rm B}$, unlike the generic phase diagram as a function of doping reported thus far \cite{Mukuda}.  Multi-layered cuprates provide a unique opportunity to research the nature of mobile holes doped into ideally flat CuO$_2$ planes. 
\begin{figure}[tpb]
\begin{center}
\includegraphics[scale = 0.39]{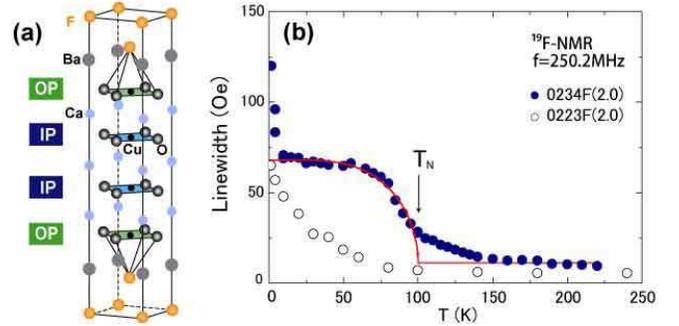}
\end{center}
\caption{\footnotesize (color online) (a)~Crystal structure of the four-layered 0234F($2.0$) with apical F$^{-1}$ ions including two IPs and two OPs~\cite{Iyo2}. All of the apical O$^{-2}$ ions are replaced by F$^{-1}$ ions. (b) The $T$ dependence of $^{19}$F-NMR spectral widths for 0234F(2.0) and 0223F(2.0) with the field perpendicular to the $c$-axis. The experimental data for 0234F(2.0) are in good agreement with the mean-field approximation for spin=1/2 with $T_{\rm N}\simeq$ 100 K as shown by solid curve, which demonstrates that a three-dimensional long-range AFM order develops for 0234F(2.0), but not for 0223F(2.0).}
\label{fig:0234F-FNMR}
\end{figure}

An other family of multi-layered HTSC systems is the four-layered cuprate Ba$_2$Ca$_3$Cu$_4$O$_8$F$_2$ with apical fluorine (F), which includes two IPs and two OPs, as shown in Fig.\ref{fig:0234F-FNMR}(a) \cite{Iyo2}. 
The remarkable feature of this cuprate is that all of the apical oxygen sites are replaced by fluorine ions (F$^{-1}$)  so that a {\it nominal Cu valence} is +2 on average. In this case, an antiferromagnetic insulating state is expected to occur, but surprisingly superconductivity with $T_{\rm c}$=55 K takes place. 
From the insights via the recent works of ARPES \cite{Chen} and band calculation \cite{OK}, Ba$_2$Ca$_3$Cu$_4$O$_8$F$_2$ appears to be the first~{\it self-doped} high-$T_{\rm c}$ superconductor having Fermi surfaces consisting of electron- and hole-doped pairs of sheets. 

In this letter, we report on the magnetic properties of the four-layered Ba$_2$Ca$_3$Cu$_4$O$_8$F$_2$ (0234F($2.0$))  and the related tri-layered cuprate Ba$_2$Ca$_2$Cu$_3$O$_6$F$_2$ (0223F(2.0)) by the site-selective NMR studies using $^{63,65}$Cu at the OP and IP and $^{19}$F at the apical site. We provide microscopic evidence that the OPs are electron-doped and hence the IPs are hole-doped as a consequence that electrons are transferred from IP into OP. We observe that the uniform mixing of AFM and HTSC occurs both at the electron-doped OP and the  hole-doped IP in 0234F($2.0$).

Polycrystalline powder samples of 0234F(2.0) and 0223F(2.0) were prepared by using a high-pressure synthesis technique as described elsewhere \cite{Iyo1,Iyo2}. The $T_{\rm c}$'s for 0234F(2.0) and 0223F(2.0) are determined to be 55 K and 76 K, respectively by the onset of diamagnetism using dc SQUID measurements. Powder X-ray diffraction measurements indicate that the samples are composed of almost a single phase. For the NMR measurements, $c$-axis oriented powder samples were used in this work.

\begin{figure}[tb]
\begin{center}
\includegraphics[width=0.8\linewidth]{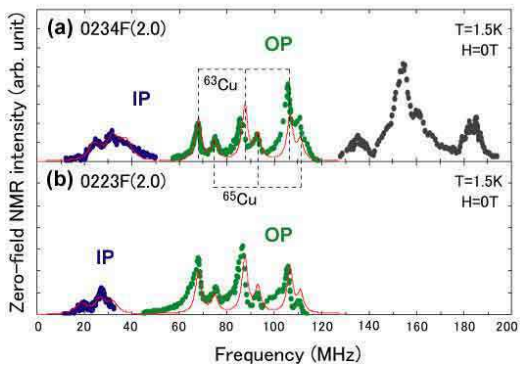}
\end{center}
\caption{\footnotesize (color online) (a)  $^{63,65}$Cu-NMR spectra of (a) 0234F(2.0) and (b) 0223F(2.0) at $T$=1.5 K and zero field. The spectra in the range 50-120~MHz with a narrow linewidth for OP are well reproduced by assuming $H_{\rm int}$=7.5 T and $^{63}\nu_{\rm Q}$(OP)=49.5 MHz for both compounds as shown by solid (red) curves. The spectra in the range 20-50~MHz for IP are simulated by assuming $^{63}\nu_{\rm Q}$(IP)=13.7 MHz, and $H_{\rm int}$=2.7 T and 2.2 T for 0234F(2.0) and 0223F(2.0), respectively. The spectra in the 130-200~MHz range arise from Cu sites in extrinsic phases affected by some disorder due to stacking faults and/or the deficiency of apical F$^{-1}$ ions. }
\label{fig:0234F(2.0)spectra}
\end{figure}

Although field-swept Cu-NMR spectra arising from the IP and OP were previously separately observed in most multilayered cuprates if they were in the paramagnetic state \cite{Tokunaga,Kotegawa2001,Kotegawa2004}, they are not observed in 0234F(2.0) because of the broad NMR spectral width due to AFM order. In fact, the onset of AFM order is evidenced by the $^{19}$F-NMR measurements on 0234F(2.0). As shown in Fig. \ref{fig:0234F-FNMR}(b), the temperature ($T$) dependence of the $^{19}$F-NMR spectral width exhibits a rapid increase below $\sim$ 100 K, followed by a saturation below $\sim 60$ K. 
As indicated by the solid (red) curve in Fig.\ref{fig:0234F-FNMR}(b), its $T$ dependence is in good agreement with the development of an AFM ordered moment expected from the mean-field approximation for spin=1/2. 
This result convinces us of a three-dimensional long-range AFM order below $T_{\rm N}\simeq$ 100 K. 
Note that the steep increase in the spectral width below $\sim$ 10 K may be ascribed to the localization of carriers in some part of this sample due to a disorder effect, which is also corroborated by the zero-field Cu-NMR experiments as described later. 
It is noteworthy that this behavior resembles that for the La-NQR spectral width which steeply increases in La$_{2-x}$Sr$_x$CuO$_4$ with $0.01 < x < 0.02$ \cite{Borsa,Ishida}. In this context, we suggest that this increase in $^{19}$F-NMR spectral width below 10 K is extrinsic in 0234F(2.0) as well. 

To further characterize the AFM order, $^{63,65}$Cu-NMR experiments have been carried out at zero field.  Magnetically ordered Cu moments induce a large internal field ($H_{\rm int}$) at the Cu sites at either IP or OP. As a result,  $^{63,65}$Cu-NMR spectra become observable at zero field, allowing us to estimate the size of Cu ordered moments. 
Actually the zero-field Cu-NMR spectra at $T=1.5$ K in Fig.\ref{fig:0234F(2.0)spectra} are well separated with many resonance peaks distributed over a wide frequency range. This result reveals that several Cu sites exist with different values of $H_{\rm int}$ and hence of Cu ordered moments at either IP or OP in this compound. 
\begin{figure}[htbp]
\begin{center}
\includegraphics[width=0.9\linewidth]{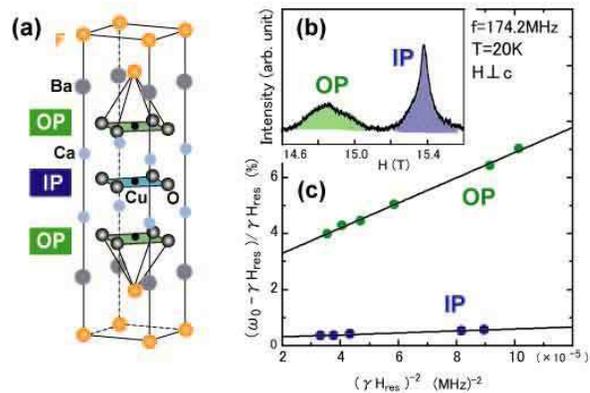}
\end{center}
\caption{\footnotesize (color online) (a) Crystal structure of the tri-layered cuprate 0223F(2.0) with apical F$^{-1}$ ions including two OPs and an IP. (b) Cu-NMR spectra of 0223F(2.0) at 20 K with the field perpendicular to the $c$-axis. The spectra from OP and IP are separately observed. (c) The $^{63}\nu_{\rm Q}$'s at OP and IP in 0223F(2.0) are evaluated to be 49.5 and 13.7 MHz, respectively from the slope of linear solid lines (see the text). }
\label{fig:0223F}
\end{figure}

In order to assign these zero-field Cu-NMR spectra to either the OP or IP in 0234F(2.0), we have investigated the related tri-layered 0223F(2.0) compound shown in Fig.\ref{fig:0223F}(a). This is because 0223F(2.0) stays in the paramagnetic state down to at least 20 K so that it enables us  to precisely determine the nuclear quadrupole frequency $\nu_{\rm Q}$ and the Knight shift for the IP and OP.  
The site assignment of Cu-NMR spectra to either IP or OP is possible from the value of $\nu_{\rm Q}$ that is proportional to the electric field gradient (EFG) at the Cu nuclear site, since $\nu_{\rm Q}$ at the OP is usually larger than that at IP.

Figure \ref{fig:0223F}(b) presents the $^{63}$Cu-NMR spectra at 20 K for 0223F(2.0) with the field perpendicular to the c-axis. Provided that the $\nu_{\rm Q}$ of $^{63,65}$Cu with nuclear spin $I$=3/2 is sufficiently smaller than the Zeeman field, the central peak for the NMR spectrum (+1/2$\Leftrightarrow$-1/2 transition) is shifted by the second order perturbation of nuclear quadrupole interaction. In the case that an external field is applied perpendicular to the $c$-axis, the following relation is valid
\begin{equation}
\frac{\omega_0-\gamma_{\rm N}H_{\rm res}}{\gamma_{\rm N}H_{\rm res}}=K_{\rm ab}+\frac{3\nu_{\rm Q}^2}{16(1+K_{\rm ab})}\frac{1}{(\gamma_{\rm N}H_{\rm res})^2}
\label{eq:shift}
\end{equation}
where $\omega_0$ is an NMR frequency,  $\gamma_{\rm N}$ is the nuclear gyromagnetic ratio of Cu nucleus, $H_{\rm res}$ is the NMR resonance field and $K_{\rm ab}$ is the Knight shift with the field perpendicular to the $c$-axis. 
From the slope of the linear lines in Fig.\ref{fig:0223F}(c) the respective $\nu_{\rm Q}$'s for two peaks are evaluated to be 49.5~MHz and 13.7~MHz, which are assigned to the OP and IP, respectively
Note that $^{63}\nu_{\rm Q}(\rm {IP})$=13.7~MHz is almost the same as those for the IPs in other multi-layered cuprates\cite{Tokunaga,Kotegawa2001,Kotegawa2004}. Most remarkably, however, $^{63}\nu_{\rm Q}(\rm {OP})$=49.5~MHz for the OP with apical F$^{-1}$ ions in 0223F(2.0) is rather larger than that for the OPs with apical O$^{-2}$. 
Noting that the $\nu_{\rm Q}$ at the OP is much larger than that at the IP in the multi-layered cuprates with apical F$^{-1}$ ions, the zero-field Cu-NMR spectra for 0234F(2.0) in the range 60-120 MHz are well described by assuming an internal field $H_{\rm int}({\rm OP})=7.5$ T, $^{63}\nu_{\rm Q}({\rm OP})=49.5$ MHz and $\theta$=$76^{\circ}$, as shown by the solid (red) curve in Fig.\ref{fig:0234F(2.0)spectra}(a).  Here $\theta$ is the angle between the principle axis of the EFG and $H_{\rm int}$.  Thus, the zero-field Cu-NMR spectra between 60-120~MHz arise from $^{63,65}$Cu at the OPs in 0234F(2.0). The $H_{\rm int}$ at the CuO$_2$ plane is expressed as $H_{\rm int}=|A_{\rm hf}|M=|A_{\rm ab}-4B|M$, where $A_{\rm ab}$ and $B$ are the on-site hyperfine field and the supertransferred hyperfine field from the four nearest-neighboring Cu sites, respectively. Here $A_{\rm ab}\sim$ 3.7 T/$\mu_{\rm B}$, $B({\rm OP})\sim$ 7.4 T/$\mu_{\rm B}$  and $B({\rm IP})\sim$ 6.1 T/$\mu_{\rm B}$ are assumed to be equivalent with those for Hg-1245 \cite{Kotegawa2004}. 
Using these values, the uniform AFM moment at the OP is estimated to be $M({\rm OP})\sim 0.29~\mu_{\rm B}$. 
We remark that the narrow NMR spectral width reveals that the $M({\rm OP})$ is homogeneous over the sample, demonstrating that an AFM metallic state occurs. As indicated in Fig.\ref{fig:0234F(2.0)spectra}(b), it should be noted that the  spectra for 0223F(2.0) observed around 60-120 MHz are similar to those for 0234F(2.0), suggesting that $M({\rm OP})\sim 0.29~\mu_{\rm B}$ for the OP in 0234F(2.0) is almost the same as that for the OP in 0223F(2.0). 
We remark that possible magnetic order in 0223F(2.0), however, remains short ranged because the $^{19}$F-NMR spectral width gradually increases, which differs from the behavior for the long-range AFM order with $T_{\rm N}=$100 K in 0234F(2.0) (see Fig.1(b)). 

Next we deal with the magnetic property of the IP.  By using $^{63}\nu_{\rm Q}({\rm IP})=13.7$~MHz, the zero-field Cu-NMR spectra observed in a range of 20-50 MHz in Figs.\ref{fig:0234F(2.0)spectra}(a) and (b) allow us to estimate $H_{\rm int}({\rm IP})=2.7$ T and 2.2 T for 0234F(2.0) and 0223F(2.0), respectively. From the fact that the Cu-NMR spectral width is significantly smaller for 0223F(2.0) than for 0234F(2.0) which exhibits the AFM with $T_{\rm N}=100$ K,
it is anticipated that the IP for 0223F(2.0) does not have any spontaneous ordered magnetic moment. This is because the carrier density $N_{\rm h}($IP)$\sim$ 0.13-0.15 at the IP for 0223F(2.0) is twice as large as that for 0234F(2.0) as discussed later, and hence the magnetic coupling via the IP between the OPs with $M({\rm OP})\sim 0.29 \mu_{\rm B}$ is too weak to develop any long-range order along the layers in 0223F(2.0). In this context, $H_{\rm int}({\rm IP})=2.2$ T at the IP in 0223F(2.0) should be attributed to the transferred hyperfine field $C_{\rm thf}$, which is the internal field at the IP induced by the moment of $\sim 0.29 \mu_{\rm B}$ at the OP. 
By taking $C_{\rm thf}$=1.1 T into account, we estimate $M({\rm IP})\sim 0.16 \mu_{\rm B}$ at the IP in 0234F(2.0) from the relation of $H_{\rm int}({\rm IP})=|A-4B|M({\rm IP})-C_{\rm thf}M({\rm OP})$.
Here we should note that the spectra between 120-200 MHz in Fig.\ref{fig:0234F(2.0)spectra}(a) resemble the zero-field Cu-NMR spectra in the disorder-driven AFM insulating states of Cu-1245(OPT) \cite{MukudaLo} and hence some disorder due to stacking faults along the $c$-axis and/or the deficiency of apical F$^{-1}$ makes carriers localize in some part of the sample. 
As a  result,  the fact that no paramagnetic NQR spectra around $^{63}\nu_{\rm Q}({\rm IP})$ and $^{63}\nu_{\rm Q}({\rm OP})$ were observed suggests that the homogeneous AFM moments of $M({\rm IP})\sim 0.16 \mu_{\rm B}$ and $M({\rm OP})\sim0.29 \mu_{\rm B}$ exist on both the Cu sites at IPs and OPs which both are metallic, respectively. 

\begin{figure}[htbp]
\begin{center}
\includegraphics[width=0.9\linewidth]{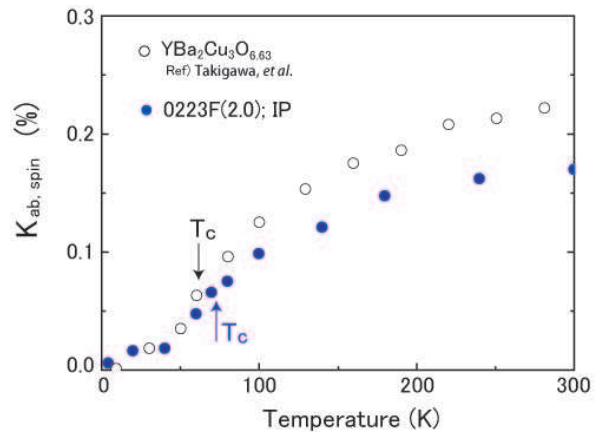}
\end{center}
\caption{\footnotesize (color online) The $T$ dependence of the spin part of the Knight shift at IP in 0223F(2.0), together with that in typical underdoped cuprate YBa$_2$Cu$_3$O$_{6.63}$ with $T_{\rm c}=62$ K \cite{Takigawa}.  }
\label{fig:0223F-Knight}
\end{figure}

\begin{figure}[tbp]
\begin{center}
\includegraphics[width=0.94\linewidth]{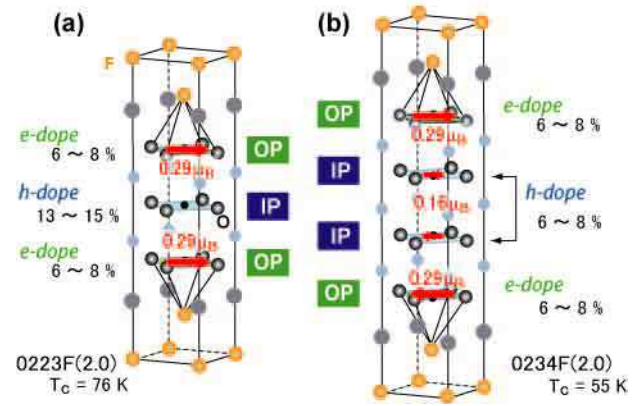}
\end{center}
\caption{\footnotesize (color online) Illustrations of magnetic properties with electron-doped OPs and hole-doped IPs for (a) 0223F(2.0) and (b)0234F(2.0). Thanks to the results from  ARPES for 0234F(2.0) which  have revealed that the Fermi surfaces consist of electron- and hole-doped sheets with their respective SC gaps \cite{Chen}, it is concluded that the uniform mixing of AFM and SC realizes both in the  hole-doped IPs and the  electron-doped OPs. }
\label{fig:electron phases}
\end{figure}

Next we address whether electrons or holes are doped into either the OP or IP and how much the carrier density is.  Figure \ref{fig:0223F-Knight} shows the $T$ dependence of $K_{\rm s}({\rm IP})$ at the IP in 0223F(2.0) with the field perpendicular to the $c$-axis.
$K_{\rm s}({\rm IP})$ decreases upon cooling, followed by a steep decrease below $T_{\rm c}$=76 K, which is characteristic for the under-doped cuprates with hole doping. In fact, it resembles that for YBa$_2$Cu$_3$O$_{6.63}$ with $T_{\rm c}$=62~K as shown by the open circles in the figure \cite{Takigawa}.  When noting that the $K_{\rm s}$(300 K) at room temperature is slightly smaller than that of YBa$_2$Cu$_3$O$_{6.63}$ with $N_{\rm h}\sim 0.18$ \cite{Cava} and $K_{\rm s}$(300 K) is empirically related to a local carrier density in a CuO$_2$ plane \cite{Kotegawa2001}, the local hole density at the IP in 0223F(2.0) is approximately estimated as  $N_{\rm h}$(IP)$\sim$ 0.13-0.15. Since the Cu valence is nominally +2 in the 0223F(2.0) on average, charge balance requires that electrons are transferred from the IP to OP. Thus, electrons are doped into each of the two OPs with approximately $N_{\rm e}$(OP)$\sim$ 0.06-0.08 because the hole doping level at the IP is anticipated to be $N_{\rm h}$(OP)$\sim$ 0.13-0.15.
According to neutron scattering measurements on electron-doped Nd$_{2-x}$Ce$_x$CuO$_4$ \cite{NdCeCuOND}, the electron-doped CuO$_2$ plane with  $N_{\rm e}\sim$0.08 is in an AFM metallic state with $M\sim 0.3 \mu_{\rm B}$. In fact, since  $M({\rm OP})\sim 0.29 \mu_{\rm B}$  is the same for the tri-layered 0223F(2.0) and the four-layered 0234F(2.0), $N_{\rm e}({\rm OP})\sim$0.06-0.08 is probably also the same for both. Therefore, $N_{\rm h}$(IP)$\sim$ 0.06-0.08 is estimated for the four-layered 0234F(2.0).  As a consequence, it is concluded that both compounds with apical F$^{-1}$ ions are {\it self-doped} high-$T_{\rm c}$ superconductors having electron doping and hole doping in one and the same compound. 
All these results make it clear that the {\it self-doping}, i.e. the transfer of electrons from the IP to OP, takes place in both compounds, as summarized in Fig.\ref{fig:electron phases}. 
We can therefore understand why $T_{\rm c}$=55 K in 0234F(2.0) increases to $T_{\rm c}$=76K in 0223F(2.0) because $N_{\rm h}$(IP)$\sim$ 0.13-0.15 for the latter is twice as large as $N_{\rm h}$(IP)$\sim$ 0.06-0.08 in 0234F(2.0). 
Further, the reason why long-range AFM order was absent in 0223F(2.0) may be because the superconducting IP with $N_{\rm h}$(IP)$\sim$ 0.13-0.15  suppresses the magnetic coupling between the OPs with $M({\rm OP})\sim 0.29 \mu_{\rm B}$.    
Here we note that our results contradict the results from the ARPES \cite{Chen} and band structure calculations \cite{OK} which suggested that electrons and holes are doped into the IP and OP, respectively, with $N_{\rm e}({\rm IP})\sim$~0.2$\pm$ 0.06 and $N_{\rm h}({\rm OP})\sim$~0.2$\pm$ 0.08. 
Our results are, however, consistent with the band calculations by Hamada who suggested that OP is electron-doped when  F$^{-1}$ is substituted for the apical O$^{-2}$ \cite{Hamada}.
Nevertheless, in combination with the result from ARPES which revealed that the Fermi surfaces consist of electron- and hole-doped sheets with the SC gap on the former sheets twice that on the latter one \cite{Chen}, the present work shows that 0234F(2.0) is a {\it self-doped} AFM high-$T_{\rm c}$ superconductor with $T_{\rm N}$=100 K and $T_{\rm c}$=55 K. 


In conclusion, $^{19}$F- and $^{63,65}$Cu-NMR studies have revealed that the four-layered 0234F(2.0) with apical fluorine F$^{-1}$ is an AFM high-temperature superconductor with $T_{\rm N}$=100 K and $T_{\rm c}$=55 K. Along with the results on the tri-layered 0223F(2.0) with $T_{\rm c}$=76 K, it has been demonstrated that electrons are transferred from the IP to OP in the  multi-layered cuprates with apical fluorine F$^{-1}$ ions, confirming on a microscopic level a new concept of ``{\it self-doping}" that was recently pointed out  by ARPES \cite{Chen}.  We remark that the OPs and the IPs in 0234F(2.0) have $M({\rm OP})$=0.29$\mu_{\rm B}$ with electron doping $N_{\rm e}({\rm OP})\sim$~0.06-0.08 and $M({\rm IP})$=0.16$\mu_{\rm B}$ with hole doping $N_{\rm h}({\rm IP})\sim$~0.06-0.08, respectively. Most notable from the present work is the fact that the uniform mixing of AFM and HTSC takes place in the under-doped regimes for both electron and hole doping as illustrated in  Fig.~\ref{fig:electron phases}(b). A remaining underlying issue is why the SC gap on the Fermi sheet with the electron-doped OP is twice as large as that with the hole-doped IP.

The authors would like to thank Y. Chen, M. Mori and T. Tohyama for their valuable discussions and comments.  This work was supported by Grant-in-Aid for Creative Scientific Research (15GS0213) from the Ministry of Education, Culture, Sports, Science and Technology (MEXT) and the 21st Century COE Program (G18) by Japan Society of the Promotion of Science (JSPS). 


\clearpage

\end{document}